\documentclass{svmult}
\usepackage{epsfig,graphicx} 
\usepackage[bottom]{footmisc}
\usepackage{natbib,url}      
\bibpunct{(}{)}{;}{a}{}{,}   
\def\cite#1{\citealp{#1}}    
\def\authorindex#1{}

\def\figspath{.}  

\def\jgr{{JGR}}

\begin{document}\newcount\preprintheader\preprintheader=1



\def\thisvolume{these proceedings}

\def\aj{{AJ}}			
\def\araa{{ARA\&A}}		
\def\apj{{ApJ}}			
\def\apjl{{ApJ}}		
\def\apjs{{ApJS}}		
\def\ao{{Appl.\ Optics}} 
\def\apss{{Ap\&SS}}		
\def\aap{{A\&A}}		
\def\aapr{{A\&A~Rev.}}		
\def\aaps{{A\&AS}}		
\def\an{{Astron.\ Nachrichten}}
\def\aspcs{{ASP Conf.\ Ser.}}
\def\azh{{AZh}}			
\def\baas{{BAAS}}		
\def\jrasc{{JRASC}}		
\def\memras{{MmRAS}}		
\def\mnras{{MNRAS}}
\def\nat{{Nat}}		
\def\pra{{Phys.\ Rev.\ A}} 
\def\prb{{Phys.\ Rev.\ B}}		
\def\prc{{Phys.\ Rev.\ C}}		
\def\prd{{Phys.\ Rev.\ D}}		
\def\prl{{Phys.\ Rev.\ Lett}}	
\def\pasp{{PASP}}
\def\pasj{{PASJ}}		
\def\qjras{{QJRAS}}
\def\science{{Sci}}		
\def\skytel{{S\&T}}		
\def\solphys{{Solar\ Phys.}} 
\def\sovast{{Soviet\ Ast.}}  
\def\ssr{{Space\ Sci.\ Rev.}}
\def\svassp{{Astrophys.\ Space Science Proc.}}
\def\zap{{ZAp}}			
\let\astap=\aap
\let\apjlett=\apjl
\let\apjsupp=\apjs

\def\ion#1#2{{\rm #1}\,{\uppercase{#2}}}  
\def\deg{\hbox{$^\circ$}}
\def\sun{\hbox{$\odot$}}
\def\earth{\hbox{$\oplus$}}
\def\la{\mathrel{\hbox{\rlap{\hbox{\lower4pt\hbox{$\sim$}}}\hbox{$<$}}}}
\def\ga{\mathrel{\hbox{\rlap{\hbox{\lower4pt\hbox{$\sim$}}}\hbox{$>$}}}}
\def\sq{\hbox{\rlap{$\sqcap$}$\sqcup$}}
\def\arcmin{\hbox{$^\prime$}}
\def\arcsec{\hbox{$^{\prime\prime}$}}
\def\fd{\hbox{$.\!\!^{\rm d}$}}
\def\fh{\hbox{$.\!\!^{\rm h}$}}
\def\fm{\hbox{$.\!\!^{\rm m}$}}
\def\fs{\hbox{$.\!\!^{\rm s}$}}
\def\fdg{\hbox{$.\!\!^\circ$}}
\def\farcm{\hbox{$.\mkern-4mu^\prime$}}
\def\farcs{\hbox{$.\!\!^{\prime\prime}$}}
\def\fp{\hbox{$.\!\!^{\scriptscriptstyle\rm p}$}}
\def\micron{\hbox{$\mu$m}}
\def\onehalf{\hbox{$\,^1\!/_2$}}	
\def\onethird{\hbox{$\,^1\!/_3$}}
\def\twothirds{\hbox{$\,^2\!/_3$}}
\def\onequarter{\hbox{$\,^1\!/_4$}}
\def\threequarters{\hbox{$\,^3\!/_4$}}
\def\ubv{\hbox{$U\!BV$}}		
\def\ubvr{\hbox{$U\!BV\!R$}}		
\def\ubvri{\hbox{$U\!BV\!RI$}}		
\def\ubvrij{\hbox{$U\!BV\!RI\!J$}}		
\def\ubvrijh{\hbox{$U\!BV\!RI\!J\!H$}}		
\def\ubvrijhk{\hbox{$U\!BV\!RI\!J\!H\!K$}}		
\def\ub{\hbox{$U\!-\!B$}}		
\def\bv{\hbox{$B\!-\!V$}}		
\def\vr{\hbox{$V\!-\!R$}}		
\def\ur{\hbox{$U\!-\!R$}}


\def\labelitemi{{\bf --}}  

\def\rmit#1{{\it #1}}              
\def\rmit#1{{\rm #1}}              
\def\etal{\rmit{et al.}}           
\def\etc{\rmit{etc.}}           
\def\ie{\rmit{i.e.,}}              
\def\eg{\rmit{e.g.,}}              
\def\cf{cf.}                       
\def\viz{\rmit{viz.}}
\def\vs{\rmit{vs.}}

\def\rot{\hbox{\rm rot}}
\def\div{\hbox{\rm div}}
\def\lesssim{\mathrel{\hbox{\rlap{\hbox{\lower4pt\hbox{$\sim$}}}\hbox{$<$}}}}
\def\gtrsim{\mathrel{\hbox{\rlap{\hbox{\lower4pt\hbox{$\sim$}}}\hbox{$>$}}}}
\def\dif{\: {\rm d}}                       
\def\ep{\:{\rm e}^}                        
\def\dash{\hbox{$\,-\,$}}                  
\def\is{\!=\!}                             

\def\starname#1#2{${#1}$\,{\rm {#2}}}  
\def\Teff{\hbox{$T_{\rm eff}$}}   

\def\kms{\hbox{km$\;$s$^{-1}$}}
\def\Mxcm{\hbox{Mx\,cm$^{-2}$}}    

\def\Bapp{\hbox{$B_{\rm app}$}}    

\def\komega{($k, \omega$)}                 
\def\kf{($k_h,f$)}                         
\def\VminI{\hbox{$V\!\!-\!\!I$}}           
\def\IminI{\hbox{$I\!\!-\!\!I$}}           
\def\VminV{\hbox{$V\!\!-\!\!V$}}           
\def\Xt{\hbox{$X\!\!-\!t$}}                

\def\level #1 #2#3#4{$#1 \: ^{#2} \mbox{#3} ^{#4}$}   

\def\specchar#1{\uppercase{#1}}    
\def\AlI{\mbox{Al\,\specchar{i}}}  
\def\BI{\mbox{B\,\specchar{i}}} 
\def\BII{\mbox{B\,\specchar{ii}}}  
\def\BaI{\mbox{Ba\,\specchar{i}}}  
\def\BaII{\mbox{Ba\,\specchar{ii}}} 
\def\CI{\mbox{C\,\specchar{i}}} 
\def\CII{\mbox{C\,\specchar{ii}}} 
\def\CIII{\mbox{C\,\specchar{iii}}} 
\def\CIV{\mbox{C\,\specchar{iv}}} 
\def\CaI{\mbox{Ca\,\specchar{i}}} 
\def\CaII{\mbox{Ca\,\specchar{ii}}} 
\def\CaIII{\mbox{Ca\,\specchar{iii}}} 
\def\CoI{\mbox{Co\,\specchar{i}}} 
\def\CrI{\mbox{Cr\,\specchar{i}}} 
\def\CriI{\mbox{Cr\,\specchar{ii}}} 
\def\CsI{\mbox{Cs\,\specchar{i}}} 
\def\CsII{\mbox{Cs\,\specchar{ii}}} 
\def\CuI{\mbox{Cu\,\specchar{i}}} 
\def\FeI{\mbox{Fe\,\specchar{i}}} 
\def\FeII{\mbox{Fe\,\specchar{ii}}} 
\def\FeIX{\mbox{Fe\,\specchar{ix}}}
\def\FeX{\mbox{Fe\,\specchar{x}}}
\def\FeXVI{\mbox{Fe\,\specchar{xvi}}}
\def\FrI{\mbox{Fr\,\specchar{i}}}
\def\HI{\mbox{H\,\specchar{i}}} 
\def\HII{\mbox{H\,\specchar{ii}}} 
\def\Hmin{\hbox{\rmH$^{^{_{\scriptstyle -}}}$}}      
\def\Hemin{\hbox{{\rm He}$^{^{_{\scriptstyle -}}}$}} 
\def\HeI{\mbox{He\,\specchar{i}}} 
\def\HeII{\mbox{He\,\specchar{ii}}} 
\def\HeIII{\mbox{He\,\specchar{iii}}} 
\def\KI{\mbox{K\,\specchar{i}}} 
\def\KII{\mbox{K\,\specchar{ii}}} 
\def\KIII{\mbox{K\,\specchar{iii}}} 
\def\LiI{\mbox{Li\,\specchar{i}}} 
\def\LiII{\mbox{Li\,\specchar{ii}}} 
\def\LiIII{\mbox{Li\,\specchar{iii}}} 
\def\MgI{\mbox{Mg\,\specchar{i}}} 
\def\MgII{\mbox{Mg\,\specchar{ii}}} 
\def\MgIII{\mbox{Mg\,\specchar{iii}}} 
\def\MnI{\mbox{Mn\,\specchar{i}}} 
\def\NI{\mbox{N\,\specchar{i}}}
\def\NaI{\mbox{Na\,\specchar{i}}}
\def\NaII{\mbox{Na\,\specchar{ii}}}
\def\NaIII{\mbox{Na\,\specchar{iii}}} 
\def\NiI{\mbox{Ni\,\specchar{i}}} 
\def\NiII{\mbox{Ni\,\specchar{ii}}}
\def\NiIII{\mbox{Ni\,\specchar{iii}}} 
\def\OI{\mbox{O\,\specchar{i}}} 
\def\OVI{\mbox{O\,\specchar{vi}}}
\def\RbI{\mbox{Rb\,\specchar{i}}} 
\def\SII{\mbox{S\,\specchar{ii}}} 
\def\SiI{\mbox{Si\,\specchar{i}}} 
\def\SiII{\mbox{Si\,\specchar{ii}}} 
\def\SrI{\mbox{Sr\,\specchar{i}}}
\def\SrII{\mbox{Sr\,\specchar{ii}}}
\def\TiI{\mbox{Ti\,\specchar{i}}} 
\def\TiII{\mbox{Ti\,\specchar{ii}}} 
\def\TiIII{\mbox{Ti\,\specchar{iii}}} 
\def\TiIV{\mbox{Ti\,\specchar{iv}}} 
\def\VI{\mbox{V\,\specchar{i}}} 
\def\HtwoO{\mbox{H$_2$O}}        
\def\Otwo{\mbox{O$_2$}}          

\def\Halpha{\mbox{H\hspace{0.1ex}$\alpha$}} 
\def\Ha{\mbox{H\hspace{0.2ex}$\alpha$}}
\def\Hbeta{\mbox{H\hspace{0.2ex}$\beta$}}
\def\Hgamma{\mbox{H\hspace{0.2ex}$\gamma$}}
\def\Hdelta{\mbox{H\hspace{0.2ex}$\delta$}}
\def\Hepsilon{\mbox{H\hspace{0.2ex}$\epsilon$}}
\def\Hzeta{\mbox{H\hspace{0.2ex}$\zeta$}}
\def\Lyalpha{\mbox{Ly$\hspace{0.2ex}\alpha$}}
\def\Lybeta{\mbox{Ly$\hspace{0.2ex}\beta$}}
\def\Lygamma{\mbox{Ly$\hspace{0.2ex}\gamma$}}
\def\Lycont{\mbox{Ly\hspace{0.2ex}{\small cont}}}
\def\Baalpha{\mbox{Ba$\hspace{0.2ex}\alpha$}}
\def\Babeta{\mbox{Ba$\hspace{0.2ex}\beta$}}
\def\Bacont{\mbox{Ba\hspace{0.2ex}{\small cont}}}
\def\Paalpha{\mbox{Pa$\hspace{0.2ex}\alpha$}}
\def\Bralpha{\mbox{Br$\hspace{0.2ex}\alpha$}}

\def\NaD{\mbox{Na\,\specchar{i}\,D}}    
\def\NaDone{\mbox{Na\,\specchar{i}\,\,D$_1$}}
\def\NaDtwo{\mbox{Na\,\specchar{i}\,\,D$_2$}}
\def\NaID{\mbox{Na\,\specchar{i}\,\,D}}
\def\NaIDone{\mbox{Na\,\specchar{i}\,\,D$_1$}}
\def\NaIDtwo{\mbox{Na\,\specchar{i}\,\,D$_2$}}
\def\Done{\mbox{D$_1$}}
\def\Dtwo{\mbox{D$_2$}}

\def\Mgbone{\mbox{Mg\,\specchar{i}\,b$_1$}}
\def\Mgbtwo{\mbox{Mg\,\specchar{i}\,b$_2$}}
\def\Mgbthree{\mbox{Mg\,\specchar{i}\,b$_3$}}
\def\MgIb{\mbox{Mg\,\specchar{i}\,b}}
\def\MgIbone{\mbox{Mg\,\specchar{i}\,b$_1$}}
\def\MgIbtwo{\mbox{Mg\,\specchar{i}\,b$_2$}}
\def\MgIbthree{\mbox{Mg\,\specchar{i}\,b$_3$}}

\def\CaIIK{\mbox{Ca\,\specchar{ii}\,K}}       
\def\CaIIH{\mbox{Ca\,\specchar{ii}\,H}}
\def\CaIIHK{\mbox{Ca\,\specchar{ii}\,H\,\&\,K}}
\def\HK{\mbox{H\,\&\,K}}
\def\Kthree{\mbox{K$_3$}}      
\def\Hthree{\mbox{H$_3$}}
\def\Ktwo{\mbox{K$_2$}}
\def\Htwo{\mbox{H$_2$}}
\def\Kone{\mbox{K$_1$}}     
\def\Hone{\mbox{H$_1$}}     
\def\KtwoV{\mbox{K$_{2V}$}}
\def\KtwoR{\mbox{K$_{2R}$}}
\def\KoneV{\mbox{K$_{1V}$}}
\def\KoneR{\mbox{K$_{1R}$}}
\def\HtwoV{\mbox{H$_{2V}$}}
\def\HtwoR{\mbox{H$_{2R}$}}
\def\HoneV{\mbox{H$_{1V}$}}
\def\HoneR{\mbox{H$_{1R}$}}

\def\hk{\mbox{h\,\&\,k}}
\def\kthree{\mbox{k$_3$}}    
\def\hthree{\mbox{h$_3$}}
\def\ktwo{\mbox{k$_2$}}
\def\htwo{\mbox{h$_2$}}
\def\kone{\mbox{k$_1$}}     
\def\hone{\mbox{h$_1$}}     
\def\ktwoV{\mbox{k$_{2V}$}}
\def\ktwoR{\mbox{k$_{2R}$}}
\def\koneV{\mbox{k$_{1V}$}}
\def\koneR{\mbox{k$_{1R}$}}
\def\htwoV{\mbox{h$_{2V}$}}
\def\htwoR{\mbox{h$_{2R}$}}
\def\honeV{\mbox{h$_{1V}$}}
\def\honeR{\mbox{h$_{1R}$}}

\title*{Evolution of Near-Sun Solar Wind Turbulence}

\author{P.K. Manoharan}
\authorindex{Manoharan, P. K.}

\institute{Radio Astronomy Centre, National Centre for Radio Astrophysics, \\
         Tata Institute of Fundamental Research, Udhagamandalam (Ooty),
         India}

\maketitle
\setcounter{footnote}{0}  

\begin{abstract}

This paper presents a preliminary analysis of the turbulence spectrum
of the solar wind in the near-Sun region $R < 50 R_\odot$, obtained
from interplanetary scintillation measurements with the Ooty Radio
Telescope at 327~MHz. The results clearly show that the scintillation
is dominated by density irregularities of size about $100-500$~km.  The
scintillation at the small-scale side of the spectrum, although
significantly less in magnitude, has a flatter spectrum than the
larger-scale dominant part. Furthermore, the spectral power contained
in the flatter portion rapidly increases closer to the Sun.  These
results on the turbulence spectrum for $R < 50 R_\odot$ quantify the
evidence for radial evolution of the small-scale fluctuations
($\leq50$~km) generated by Alfv\'en waves.

\end{abstract}

\section{Introduction}

The solar wind is highly variable and inhomogeneous, and exhibits
fluctuations over a wide range of spatial and temporal scales. The
properties of these fluctuations, as they move outward in the 
solar corona, are controlled by the presence of both waves and 
turbulence (\eg\ \cite{manoharan-1968ApJ...153..371C}, 
\cite{manoharan-1971JGR....76.3534B}). However, 
their relative contributions to the heating and acceleration of the
solar wind have yet to be assessed fully
(\cite{manoharan-1995SSRv...73....1T},
\cite{manoharan-2005JGRA..11003101H}). 

Radio scattering and scintillation experiments measure density 
fluctuations,
which are related to the wave field, density fluctuations, and magnetic 
turbulence (\eg\ \cite{manoharan-1986ApJ...309..342H}, 
\cite{manoharan-1987JGR....92..282M}). 
The density fluctuation spectrum roughly follows a Kolmogorov power law in
the spatial scale range $100-1000$~km, at distances well
outside the solar wind acceleration region.  However, nearer to the Sun 
the spectrum tends to be flat (\eg\
\cite{manoharan-1979JGR....84.7288W}). The spectrum of the high-speed
streams from coronal holes is steeper than Kolmogorov decay, which
is attributed to dissipation at scales above 100~km (\eg\
\cite{{manoharan-1994JGR....9923411M},
{manoharan-2000ApJ...530.1061M}}).  There is considerable interest
to understand the radial change of the fluctuations due to both
waves and turbulence in the solar wind acceleration region.  In this
study, spectral features are analysed over a range of distances from
the Sun using interplanetary scintillation measurements made with
the Ooty Radio Telescope at 327~MHz (\cite{manoharan-1971Natur.230..185S}).

\begin{figure}
\centering
\includegraphics[width=8cm]{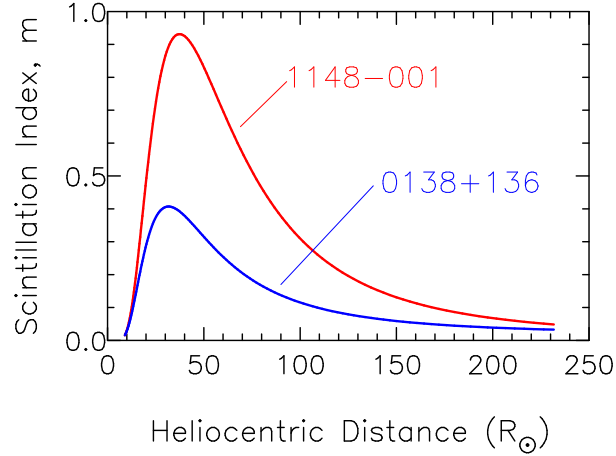}
\caption[]{\label{manoharan-fig:fig1}
  Average scintillation index against heliocentric distance observed
  at 327~MHz.  Radio quasars 1148-001 and 0138+136
  have equivalent angular diameters of about 15 and 
  50~milli-arcsec, respectively. }
\end{figure}

\section{Interplanetary scintillation}

Interplanetary scintillation (IPS) is the variability of distant
compact radio source (\eg\ a quasar or a radio galaxy) caused by
microturbulence in the solar wind of spatial scales 10 to 1000~km
(\eg\ \cite{manoharan-1994JGR....9923411M}). Scintillation
measurements normally refer to the instantaneous departure of
intensity ($\delta I(t)$) from the mean intensity of the source
($\left<I\right>$), \ie\ $\delta I(t)$ = $I(t) - \left<I\right>$.
Since the irregularities are convected by the solar wind, the
statistical fluctuations of $\delta I(t)$ can be used to estimate the
speed and turbulence spectrum of the solar wind, integrated along the
line of sight to the radio source.  However, for a given line of
sight, the spectrum of scintillation drops rapidly with distance from
the Sun, $C_{\rm N}^{2}(R)$\,\,$\sim$\,\,$R^{-4}$, and the scattering
is therefore concentrated where the line-of-sight is closest to the
Sun. The shape of the turbulence spectrum can be inferred from the
temporal IPS spectrum, obtained by taking the Fourier transformation
of intensity time series.  The rms intensity variation
$\left<{\delta I(t)^2}\right>^{1/2}$ is the integral of the power
spectrum.  The scintillation index, $m$, is estimated by
\begin{equation}
  m^2  =  \frac{1}{\left<I\right>^2} \int_{0}^{f_{\rm c}} P(f)\,\D f ,
\end{equation}
where $f_{\rm c}$ is the cutoff frequency of the spectrum at which
the scintillation equals the noise level. The systematic radial 
variation of $C_{\rm N}^{2}(R)$ can be obtained from the 
index versus distance ($m-R$) plots as in Fig.~\ref{manoharan-fig:fig1}. 
These smoothed plots represent average scintillations observed over
several years for two well-known radio quasars
(\cite{manoharan-2008psa..book..235M}).

At given heliocentric distance, a compact source scintillates more
than an extended one, because Fresnel filtering plays a key role in
producing the intensity fluctuations and the scintillation is heavily
attenuated by a large angular size $\Theta$ $\geq$ $\sqrt{\lambda/Z}$,
where $\lambda$ is the wavelength of observation and {\it Z} is the
distance to the scattering screen.  The observations reported in this
study have been made with the Ooty Radio Telescope (ORT), which
operates at $\lambda$=0.92~m.
In the case of near-Sun IPS measurements, the scattering medium is
located at about 1~AU and therefore sources having angular size
$\Theta\,>\, 500$~milli-arcsec do not scintillate.

\begin{figure}
\centering
\includegraphics[width=5.7cm,angle=-90]{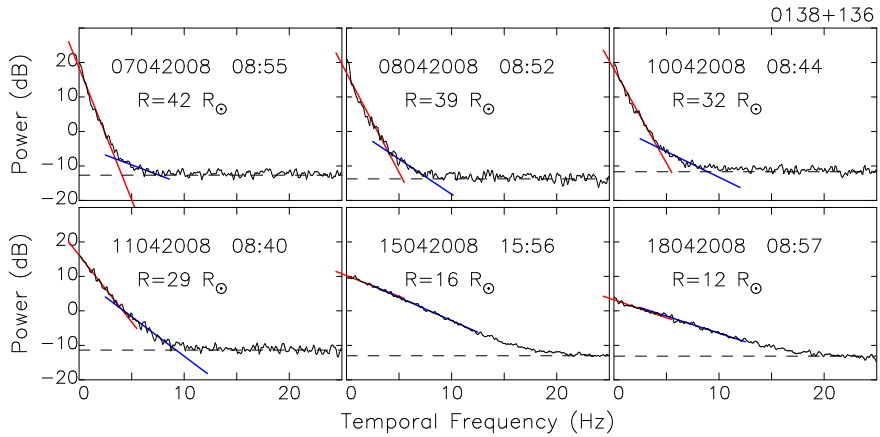}
\caption[]{\label{manoharan-fig:fig2}
  Sample temporal power spectra of 0138+136 on log-linear scale,
  showing spectral shape variations with distance from the Sun. The
  date and time of observation and the heliocentric distance ($R$) are
  specified.  These observations have been made at the eastern limb of the
  Sun so that the source approaches the Sun with increasing day
  number.}
\end{figure}

Figure~\ref{manoharan-fig:fig1} shows that as the Sun is approached, 
the scintillation increases to a peak value at a distance of 
$R \approx 40 R_\odot$, and then decreases for further closer solar 
offsets (\eg\ \cite{manoharan-1993SoPh..148..153M}),  where 1 solar 
radius is $R_\odot = 6.96 \times 10^5$~km. The peak or transition
distance, $R \approx 40 R_\odot$, is the characteristic of IPS
measurement at $\lambda$=0.92~m. It is a function of observing 
wavelength and moves close to the Sun with decreasing wavelength 
(\eg\ \cite{manoharan-1978SSRv...21..411C},
\cite{manoharan-1982RaSc...17..664K}).  At $R < 40 R_\odot$,
the scintillation index of an ideal point source saturates at $m
\approx 1$.  In the weak-scattering region with $R \geq 40 R_\odot$,
the decline in scintillation corresponds to a fall of turbulence
approximately as $1/R^2$. The shape of the temporal spectrum of
the intensity fluctuations, $P(f)$, is linearly related to the spatial
turbulence spectrum, $\Phi_{\rm N_e}(q)$, which can be obtained by
fitting the measured scintillation spectrum
(\cite{manoharan-1990MNRAS.244..691M},
\cite{manoharan-1994JGR....9923411M}).  The turbulence spectrum is
given by the power-law form $\Phi_{\rm N_e}(q) \sim q^{-\alpha}$, with
the power index $\alpha$ varying between 3 and 3.8, depending on the
solar-wind source location on the Sun.  The spatial spectrum
also includes Gaussian cutoffs at high spatial wavenumber $q$,
set by the source-size visibility and the inner (dissipation)
scale of the solar wind turbulence.

\section{Scintillation spectrum near the Sun}

\begin{figure}
\centering
\includegraphics[width=5.7cm,angle=-90]{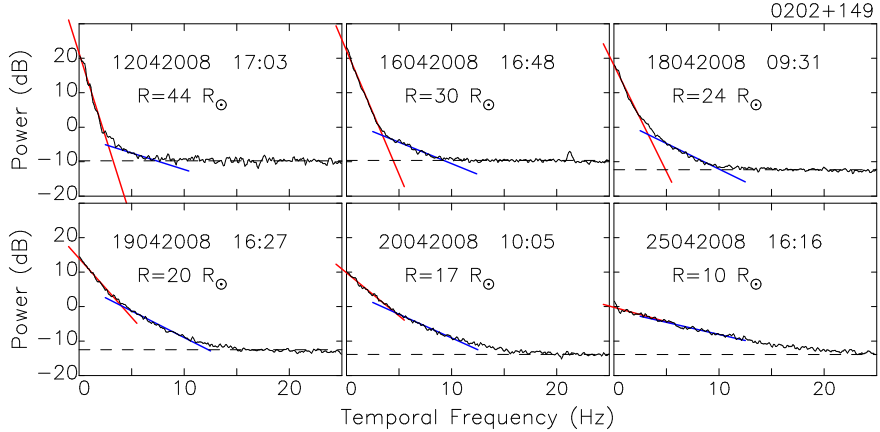}
\caption[]{\label{manoharan-fig:fig3}
  Same as Fig.~\ref{manoharan-fig:fig2}, for radio source 0202+149.}
\end{figure}

In the strong-scattering region with $R < 40 R_\odot$, the
turbulence is too intense and the relationship between the
scintillation index and $C_{\rm N}^{2}(R)$ takes a complex form.  The
decrease in scintillation is due to the smearing of scintillation
caused by the angular size of the radio source.  In a physical sense,
many independent random secondary sources exist inside the Fresnel
zone, reduce the coherence, and cause reduction of the intensity
fluctuations. However, the strong-scintillation spectrum is wider than
the weak case, and contains information on a large range of turbulence
scales.  The present study reports a systematic analysis of intensity
scintillation measurements of compact radio sources ($\Theta
\approx 50$~milli-arcsec) made with the ORT at 327~MHz.  It allows to
probe the solar wind from about 10 to 250$R_\odot$. The
closest solar offset is mainly limited by the ORT beam width.
Figures~\ref{manoharan-fig:fig2} and \ref{manoharan-fig:fig3} display 
temporal scintillation spectra of radio quasars
0138+136 and 0202+149, observed at different solar offsets during
April 2008.  The sampling rate, 50~Hz, employed in the present study
in principle extends the temporal frequency range of the spectrum to
25~Hz, which allows to infer the statistics of even small-scale
turbulence. For example, for a typical value of the solar wind speed $V$,
the spectrum can cover spatial wavenumbers in the range
$0.002<q = (2 \pi f / V) < 0.2$~km$^{-1}$, corresponding to scales in
the range 5 to 500~km.

Nearer to the Sun the spectrum broadens, suggesting 
systematic increase in turbulence associated with small-scale
irregularity structures ($< 100$~km). The flattening of the spectrum at
$R < 40 R_\odot$ indicates addition of small-scale
turbulence.  The remarkable change is that the high-frequency part of
the spectrum gradually extends into the low-frequency part at
distances closer than the transition point ($R < 40 R_\odot$).
The diminishing of spectral power at scales close to the Fresnel
radius suggests the possibility of dominant effect of Fresnel filter,
which can smear the scintillation.  At $R > 40 R_\odot$, the
low-frequency part of the spectrum gradually steepens and merges with
the slope of the density turbulence spectrum at scales smaller than
the Fresnel radius.  When a large number of spectra on a given radio
source, observed on consecutive days over a period of 45 days, are
displayed in movie mode this gives a direct visualization, making the
above results immediately apparent.

\section{Radial evolution of small- and large-scale turbulence}

\begin{figure}
\centering
\includegraphics[width=7.0cm,angle=-90]{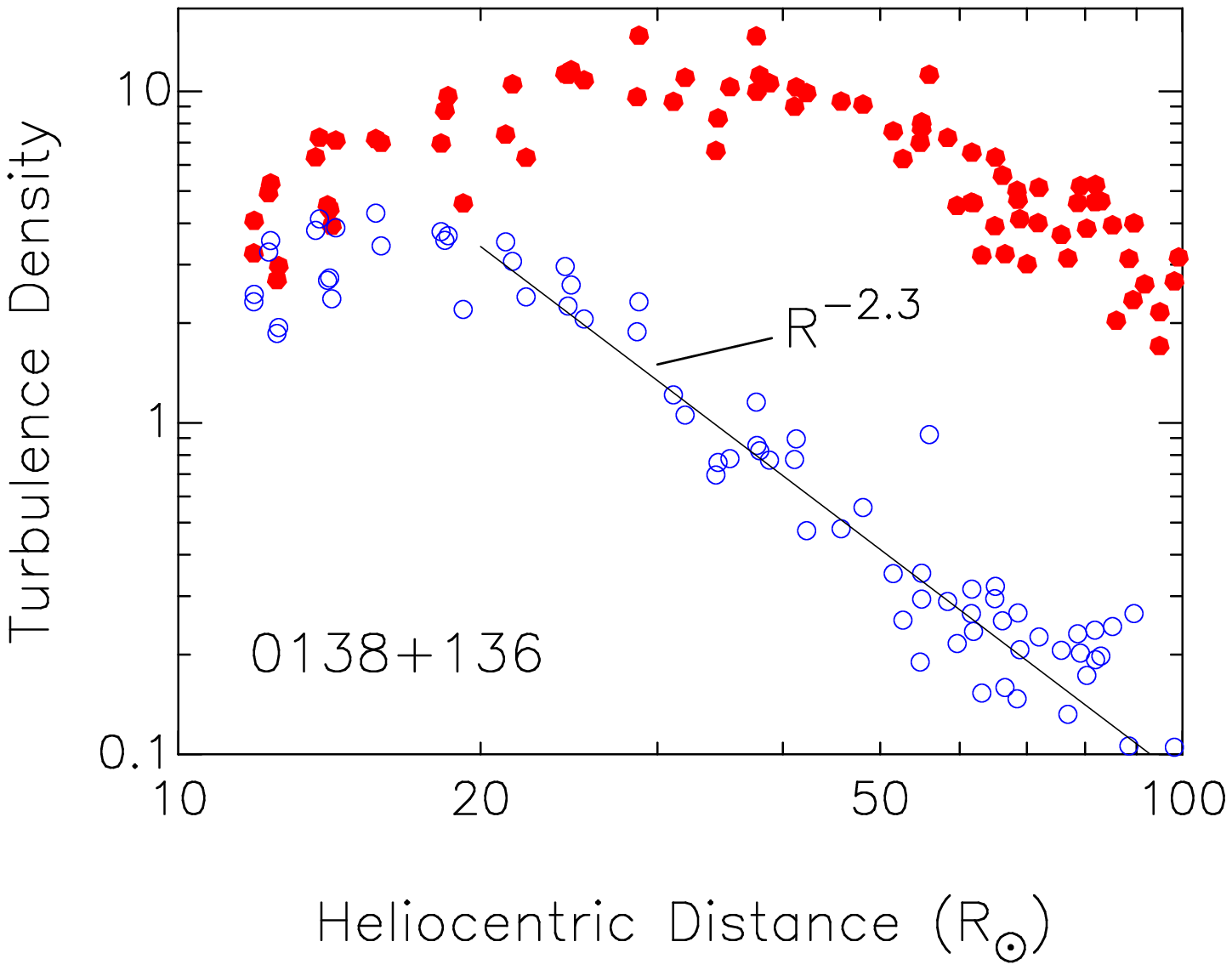}
\includegraphics[width=7.0cm,angle=-90]{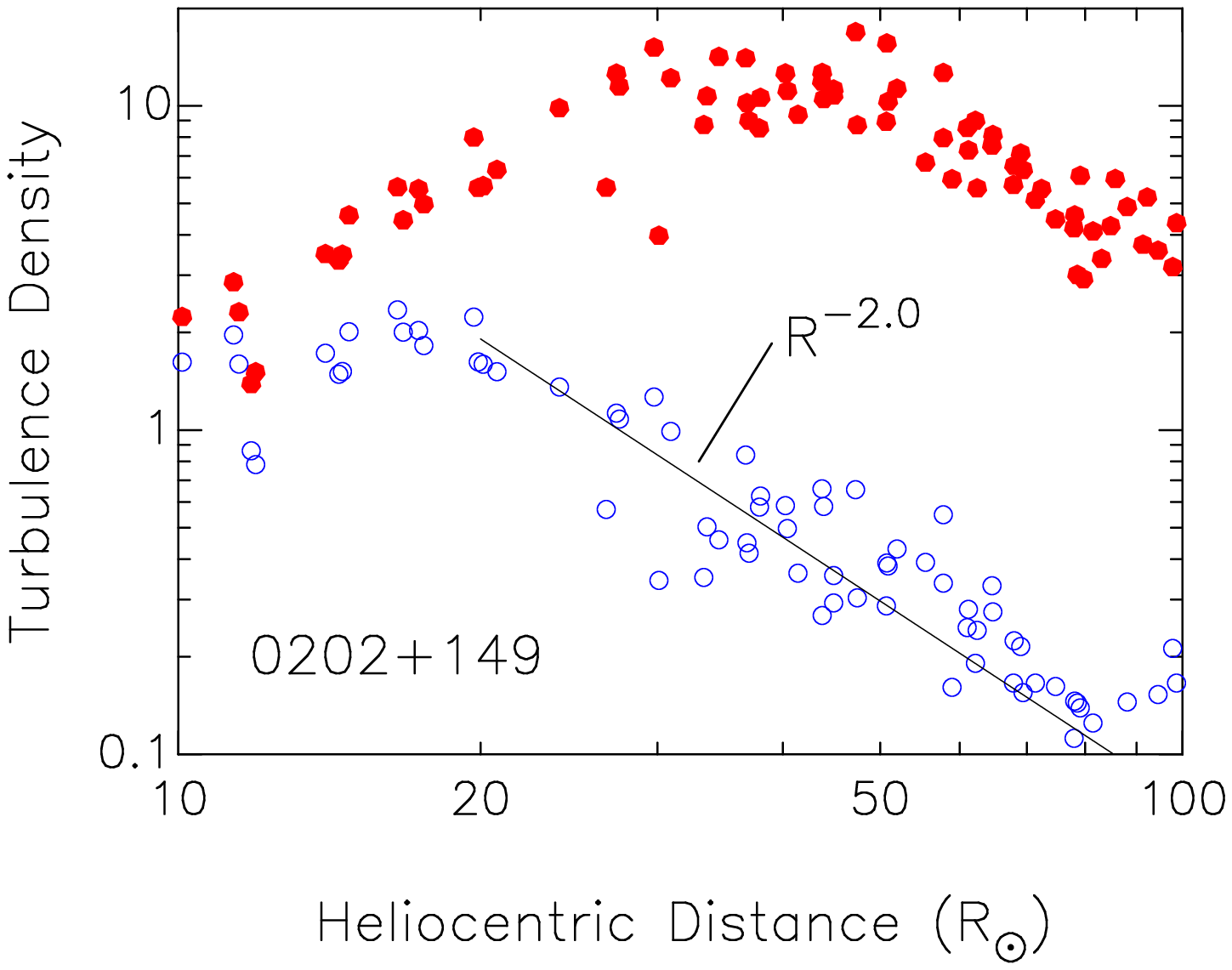}
\caption[]{\label{manoharan-fig:fig4a}
  Radial turbulence evolution in the low-frequency ({\it
  filled symbols}) and high-frequency portions ({\it open symbols}) of
  the spectrum.}
\end{figure}

Figure~\ref{manoharan-fig:fig4a} shows the power of turbulence associated with the low- and
high-frequency portions of the spectrum at different solar offsets for
radio sources 0138+136 and 0202+149. These plots illustrate the
attenuation and enhancement of the scintillations, respectively, for
the large-scale ($> 100$~km) and the small-scale ($<100$~km) spectral
regions and their radial variations.  For most of the temporal
spectra, the slope change from high- to low-frequency part is
apparent, and whenever the spectrum monotonously increases towards the
low frequency part, the half-value of the cutoff frequency (\ie\
$f_{\rm c}/2$) is considered to mark the separation of the
scintillation between the low- and high-frequency parts.

It is obvious that the turbulence density associated with the 
low-frequency part is dominant at all heliocentric distances
and that it closely follows the shape of the overall scintillation 
index versus distance curve (Fig.~\ref{manoharan-fig:fig1}). 
\citet{manoharan-1993SoPh..148..153M} has 
shown that the scintillation variation at $R > 40 R_\odot$ is of 
power-law form, with $m \sim R^{-\beta}$ and $\beta = 1.7\pm0.2$.
When the integration is accounted for, the scattering power 
changes as $C_{\rm N}^{2}(R) \sim R^{-({2\beta+1})} = R^{-4.4\pm0.4}$.
The scintillation in the low-frequency part of the spectrum is
consistent with the above radial evolution.  However, in the
high-frequency part, the scintillation increases with decreasing solar
offsets and tends to merge with the above portion. In the distance
range $R = 15 - 100 R_\odot$, the scintillation due to the 
high-frequency part follows the power-law $m_{\rm high\_freq} \sim
R^{-b}$.  Both sources show similar slopes $b \approx 
2.0$ and 2.3.  However, the average radial trend is much steeper,
with $C_{\rm N_{high\_freq}}^2(R) \sim R^{-({2b+1})} = R^{-5.3}$,
than the density turbulence slope $\sim R^{-4}$. 
The turbulence associated with small-scale fluctuations 
($\leq 50$~km) in the solar wind acceleration region steeply 
increases towards the Sun.

The strong scintillation spectra of selected radio sources observed at
Ooty have also been compared with same-day observations at higher
observing frequencies, for which the measurements fall in the weak
scintillation regime. For example, IPS measurements with the Giant
Metrewave Radio Telescope (GMRT) at 610~MHz (Manoharan et al., in
preparation) and the European Incoherent Scatter (EISCAT) system at
931~MHz (R.A. Fallows, private communication) reveal spectral features
that are closely similar to that observed with the ORT at low and high
temporal frequencies.  The contributions of low and high spatial
scales of the scintillation spectrum are not altered in the transition
from strong to weak scintillation.

\section{Discussion and conclusion }

Several IPS experiments have shown the turbulence spectrum to be
$\Phi_{\rm N_e} \sim q^{-\alpha}$, with the dissipative scale 
(\ie\ inner scale or cutoff scale) size
increasing linearly with distance as $l_{\rm i} \approx\
(R/R_\odot)^{1 \pm 0.1}$~km at $R \leq 100 R_\odot$
(\cite{manoharan-1987sowi.conf...55M},
\cite{manoharan-1989ApJ...337.1023C},
\cite{manoharan-1994JGR....9923411M}).
Further, a flatter spectrum ($\alpha \leq 3$) 
and smaller dissipative scales ($l_{\rm i} < 10$~km) have been observed 
in the near-Sun solar wind acceleration region ($R < 20 R_\odot$)
(\cite{manoharan-1989ApJ...337.1023C}, \cite{manoharan-1980SvA....24..454Y}, 
\cite{manoharan-1998JGR...103.6571Y}). 
The present result of flatter spectrum for the low-frequency part,
which at larger distances merges with the density turbulence spectrum,
is consistent with the earlier findings.

The effects of the angular structure of the radio source and of 
inner-scale turbulence dissipation are three-dimensionally Gaussian 
in shape, and tend to attenuate the high-frequency tail of the 
spectrum (\cite{manoharan-1987sowi.conf...55M}, 
\cite{manoharan-1998JGR...103.6571Y}). 
The inner-scale contribution is not significant at 
small solar offsets (it decreases and becomes small at regions
close to the Sun). Furthermore, the effect of the angular size of the 
compact radio source ($\Theta \approx 50$~milli-arcsec) is considerably 
small. However, the key point is that, in the near-Sun regions 
($R < 40 R_\odot$), a significant enhancement in scintillation
power is measured at the high-frequency portion of the spectrum, 
well above the dissipation and source-size cutoff levels.
In order to show scintillation above these cutoffs in the tail part of
the spectrum, strong fluctuations are likely to be present which are
oriented in different directions than the radial flow of the
solar wind. Therefore, the systematic and significant increase in
power at the small-scale part of the spectrum suggests an active role of
irregularities produced by magnetosonic waves in the solar wind, with
multiple scale sizes and vector directions.  The rapid radial change
of the turbulence associated with the small-scale irregularities, $C_{\rm
N(high\_freq)}^{2}(R) \sim R^{-5.3}$, indicates that the dominant
contribution is due to wave-generated turbulence in the solar wind
acceleration region.  Therefore, the overall near-Sun turbulence
spectrum can be explained by the combined effects of the smeared
density turbulence spectrum and the strong fluctuations generated by
Alfv\'en waves at small scales ($\leq 50$~km).

In summary, this preliminary analysis of the temporal spectrum of
scintillations measured in the solar wind acceleration region provides
evidence that, apart from density turbulence, small-scale fluctuations
produced by magnetosonic waves plays a key role in shaping the
spectrum.  In comparison with the density turbulence, the effect of
waves is significant but its importance decreases rather steeply with
heliocentric distance.  Its presence in the solar wind extends outside
the acceleration region ($R > 20 R_\odot$), although weaker in
intensity.  A more rigorous study of the small-scale microturbulence,
its variation with the solar cycle and solar source regions will be
reported in more detail elsewhere.

\begin{acknowledgement}
  The author thanks the observing/engineering team and research
  students of the Radio Astronomy Centre for help in performing the
  observations and the preliminary data reduction. This work is
  partially supported by the CAWSES--India Program, which is sponsored
  by the Indian Space Research Organisation (ISRO).
\end{acknowledgement}

\begin{small}

\bibliographystyle{rr-assp-bib}       


\end{small}
\end{document}